\documentclass[letterpaper,prb,twocolumn,showpacs]{revtex4}
\usepackage{graphicx,bm}
\begin{document}

\title{Spatial trends of non-collinear exchange coupling mediated
by itinerant carriers with different Fermi surfaces}

\author{Hsin-Hua Lai}
\affiliation{Physics Department, California Institute of Technology,
Pasadena, CA 91125, U.S.A.}

\author{Wen-Min Huang}
\affiliation{Department of Physics, National Tsing Hua University,
300 Hsinchu, Taiwan}

\author{Hsiu-Hau Lin}
\affiliation{Department of Physics, National Tsing-Hua University, Hsinchu 300, Taiwan}
\affiliation{Physics Division, National Center for Theoretical Sciences, Hsinchu 300, Taiwan}

\date{\today}


\begin{abstract}
We study the exchange coupling mediated by itinerant carriers with spin-orbit interaction by both analytic and numeric approaches. The mediated exchange coupling is non-collinear and its spatial trends depend on the Fermi surface topology of the itinerant carriers. Taking Rashba interaction as an example, the exchange coupling is similar to the conventional Ruderman-Kittel-Kasuya-Yosida type in weak coupling. On the other hand, in the strong coupling, the spiral interaction dominates. In addition, inclusion of finite spin relaxation always makes the non-collinear spiral exchange interaction dominant. Potential applications of our findings are explained and discussed.
\end{abstract}

\maketitle

\section{Introduction}

The central theme of spintronics is to manipulate the extra spin degrees of freedom in condensed matter systems\cite{Wolf01,Zutic04,Sun04,Sun06,Sharma,MacDonald05,Awschalom}. One of the proposals is to manipulate the spin polarization of electrons by the spin-orbit (SO) interaction\cite{Zutic04,Datta89} such as the Rashba interaction from the structural inversion asymmetry at a surface or interface in semiconductors\cite{Rashba60}. To make these proposals feasible in spintronics, it is important to achieve a flexible tuning of the interaction strength. So far, many surprising works have demonstrated how to enhance the strength of the Rashba interaction either in a semiconductor or in some metal spin-splitting surface states. For instance, by doping Bi into GaAs, it was shown that the spin-splitting gap is significantly enhanced in the alloy ${\rm GaAs_{1-x}Bi_{x}}$\cite{Francoeur,Fluegel}, which opens up an alternative approach to enhance the Rashba coupling in semiconductors. Furthermore, it was shown recently that the surface states of Bi/Ag show a giant Rashba-type spin splitting due to the surface-potential barrier\cite{Ast07}. In addition, it was also demonstrated that, by doping Pb into Bi/Ag, the Fermi energy of the surface states is tunable\cite{Ast08}. These exciting discoveries open up the possibilities to manipulate the Rashba coupling over a wide range of interaction strengths. Therefore, it is important and exciting to explore the related physics properties in both weak and strong regimes where the Fermi surface (FS) topologies are different.

Note that the Rashba interaction in a two-dimensional electron gas gives rise to two different FS topologies that may play a significant role in dictating the response functions.  The classical Dyakonov-Perel (DP) mechanism of spin relaxation\cite{DP} due to Rashba coupling shows different signatures when the FS topology changes. In weak coupling, the DP relaxation mechanism leads to exponential decay\cite{DP}. But, in strong coupling (or equivalently, the low-density regime), the DP relaxation mechanism changes into a peculiar power law\cite{Grimaldi05}. Furthermore, in strong coupling, the cancellation of the spin-Hall conductivity for spin-conserving momentum scattering\cite{Rashba04,Dimitrova} is not related to the vanishing of the vertex function, as happened in the weak regime. Instead, the cancellation is between the on-Fermi surface and off-Fermi surface contributions\cite{Grimaldi06}. Another example is the disorder-induced localization due to the Rashba coupling\cite{Chaplik}. For a two-dimensional electron gas with strong Rashba coupling, it behaves effectively as an one-dimensional system\cite{Galstyan}, which results in infinite bound states even for a short-range potential\cite{Chaplik,Grimaldi08}. The topological change of the FS also plays an interesting role in superconductivity\cite{Cappelluti}. It was shown that the critical temperature of the superconductor can be tuned by the Rashba coupling, suggesting materials with stronger Rashba coupling are good candidates for enhanced superconductivity\cite{Cappelluti}. Finally, in our previous work\cite{Huang06}, we demonstrated that the exchange coupling mediated by itinerant carriers with Rashba coupling is strongly tied up with the FS topology. We showed numerically that the non-collinear exchange coupling reveals two spatial trends depending on the FS topologies -- one is the RKKY-like oscillatory behavior and the other is the spiral.

Here we extend our previous work\cite{Huang06} by a complementary analytic approach within the path-integral formalism. This method is particularly powerful in describing the asymptotic behavior of the mediated exchange coupling and provides a comprehensive understanding of the general trends. In the weak Rashba regime, the exchange coupling consists of the RKKY oscillations at shorter length scale and a spiral background\cite{Imamura04}. On the other hand, the RKKY interaction is suppressed when entering the strong coupling regime. It is rather nice that the analytic results agree with previous numerics. We also use a phenomenological  approach to investigate the effect of the finite spin relaxation in the itinerant carriers. Since the finite relaxation rate breaks the time-reversal symmetry, it leads to the suppression of the RKKY interaction. Thus, the non-collinear spiral interaction stands out and becomes dominant.

The rest of the paper is organized as follows. In Sec. II, we revisit the ferromagnetic/normal/ferromagnetic trilayer junction mediated by a two-dimensional electron gas with Rashba interaction. The band structure and the Fermi surface topologies at different carrier concentrations are explained in detail. In Sec. III, we derive the mediated exchange coupling within linear response theory. After integrating out the angular part, the resultant formula can be implemented in numerics to calculate the effective exchange coupling. In Sec. IV, we use a complimentary analytic approach to derive the effective exchange coupling. In Sec. V, the effects of finite spin relaxation are discussed in both weak and strong Rashba regimes, followed by a brief conclusion.

\begin{figure}
\begin{center}
\includegraphics[width=7cm]{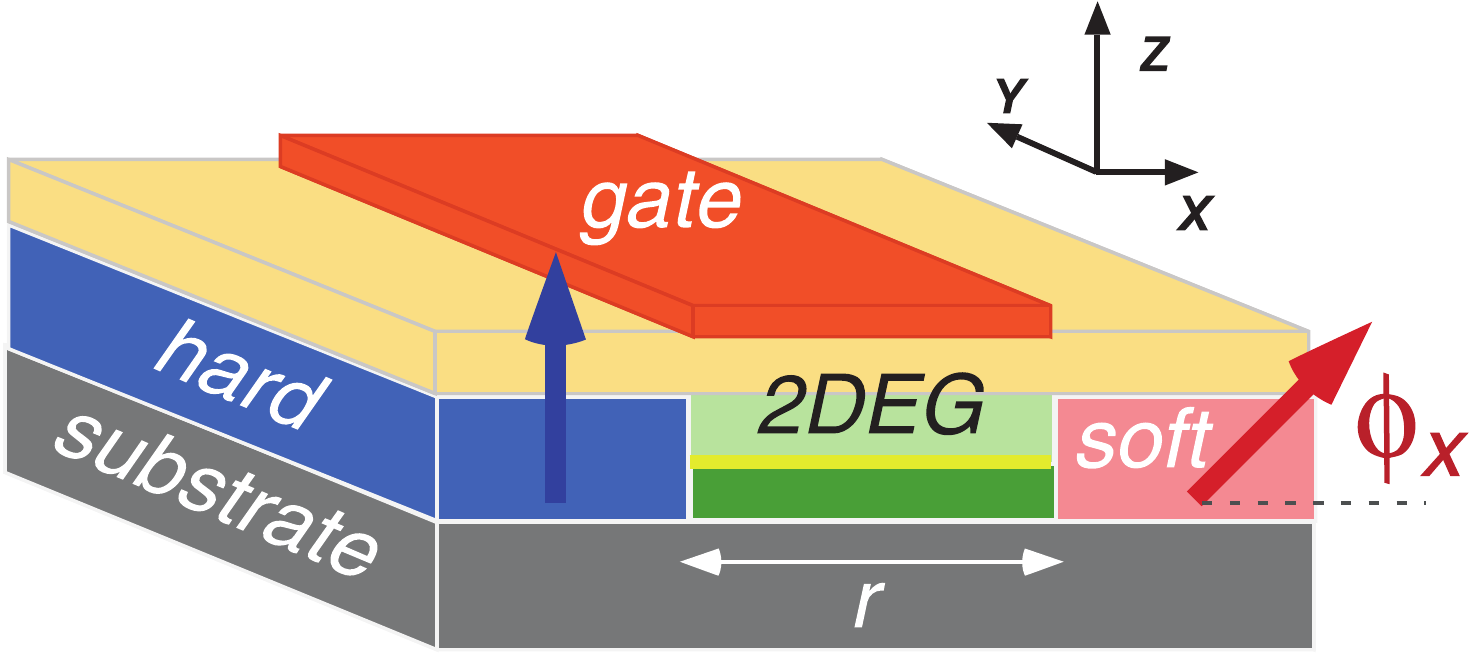}
\caption{Noncollinear exchange coupling mediated by itinerant carriers with Rashba interaction and finite spin relaxation. The hard magnet on the left is pined while the direction of the soft magnet on the right is determined by the effective exchange coupling.} \label{junction}
\end{center}
\end{figure}

\section{Linear Response Theory}

Consider a ferromagnetic/normal/ferromagnetic (F/N/F) trilayer magnetic junction (TMJ) as shown in Fig. \ref{junction}, where the intermediate layer contains a two dimensional electron gas (2DEG) with the Rashba coupling. The 2DEG in the intermediate layer is described by the Hamiltonian\cite{Rashba60},
\begin{equation}
H_R= \int d^{2}{\textbf r}  \hspace{0.2cm}\Psi^{\dagger} \left[ \frac{-\nabla^2}{2m^{*}} \textbf{1} -i\gamma_R\left(\nabla_y\sigma_x-\nabla_x\sigma_y\right)\right] \Psi,
\end{equation}
where $\gamma_R$ is the strength of the Rashba interaction and $m^*$ is the effective mass of the itinerant carriers. The two-component spinors $\Psi^{\dagger}$, $\Psi$ are the creation/annihilation operators for the itinerant carriers. After integrating out the itinerant carriers\cite{Lin04,Lin06,Huang06,Huang08}, the exchange coupling between the ferromagnetic layers is described by an effective Heisenberg Hamiltonian, $H_{\rm eff} = \sum_{ij} J_{ij} S^{i}_{L}S^{j}_{R}$. Within the linear response theory\cite{Flensberg,Mahan}, the mediated exchange coupling $J_{ij}$ is proportional to the static spin susceptibility tensor,\cite{Lin04,Lin06,Huang06,Huang08}
\begin{eqnarray}
\chi_{ij}(r,\theta)\! = \hspace{-1.5mm} \left[
\begin{array}{ccc}
g_0+g_2\cos2\theta & g_2\sin2\theta & g_1\cos\theta \\
g_2\sin2\theta & g_0-g_2\cos2\theta & g_1\sin\theta \\
-g_1\cos\theta & -g_1\sin\theta & h_0 \end{array} \right]\!.
\end{eqnarray}
It is rather remarkable that the symmetry arguments make the angular dependence explicit and reduce the numerical task down to evaluation of four real scalar functions, $g_0(r)$, $g_1(r)$, $g_2(r)$ and $h_0(r)$. After integrating out the angular parts, the first three functions can be casted into the following integral forms,
\begin{eqnarray}
g_{a}(r)\hspace{-0.05cm}=\hspace{-0.15cm}\int\hspace{-0.15cm}
\frac{d^2k_1}{2\pi}\hspace{-0.15cm}\int\hspace{-0.15cm}\frac{d^2k_2}{2\pi}
\hspace{-0.1cm}\sum_{\lambda_{1}\lambda_{2}}I^{\lambda_{1}\lambda_{2}}_{a}\hspace{-0.1cm}\left[\frac{f(\epsilon_{k_{1}\lambda_{1}})-f(\epsilon_{k_{2}\lambda_{2}})}{\epsilon_{k_{2}\lambda_{2}}-\epsilon_{k_{1}\lambda_{1}}-i\eta}\right]\hspace{-0.1cm},
\end{eqnarray}
where the dispersion for the particle with momentum $k$ and chirality $\lambda=\pm 1$ is $\epsilon_{k\lambda} = (k-\lambda k_R)^2/(2m^*)- E_R$. Here we introduce the Rashba momentum $k_R=m^*\gamma_R$ and the Rashba energy $E_R=k_R^2/2m^*$.
The oscillatory kernels (after integrating out the angular parts) in the integral are
\begin{eqnarray}
&&\hspace{-0.5cm}I^{\lambda_{1}\lambda_{2}}_{0}=J_0(k_1r)J_0(k_2r),\\
&&\hspace{-0.5cm}I^{\lambda_{1}\lambda_{2}}_{1}=-\lambda_{1}J_1(k_1r)J_0(k_2r)-\lambda_{2}J_0(k_1r)J_1(k_2r),\\
&&\hspace{-0.5cm}I^{\lambda_{1}\lambda_{2}}_{2}=-\lambda_{1}\lambda_{2}J_1(k_1r)J_1(k_2r),
\end{eqnarray}
where $J_n(x)$ is denoted the Bessel function of the first kind. For the Rashba Hamiltonian, its simpleness gives rise to an extra relation $h_0(r) = g_0(r)+g_2(r)$ beyond the symmetry argument. As a result, we only need to evaluate three independent scalar functions.

Supposing the ferromagnet on the left of the trilayer magnetic junction is aligned along the $z$-axis, we are interested in the mediated exchange coupling proportional to $\chi_{iz} (r, \theta=0)$, where $r$ is the width of intermediate layer. The induced moment is captured by the spiral angle (shown in Fig. \ref{junction}), which can be expressed in terms of $\chi_{xz}$ and $\chi_{zz}$, 
\begin{eqnarray}
\cos\left[\phi_x(r)\right] &=& \frac{\chi_{xz}(r,0)}{\sqrt{\chi^2_{xz}(r,0)+\chi^2_{zz}(r,0)}},\\
\sin\left[\phi_x(r)\right] &=& \frac{\chi_{zz}(r,0)}{\sqrt{\chi^2_{xz}(r,0)+\chi^2_{zz}(r,0)}}.
\end{eqnarray}
Therefore, by evaluating $g_a(r)$ numerically, we can study the spatial trends of the spiral angle $\phi_x(r)$.

In our numerical computations, we will change the parameter $k_R/k_F$, where $k_{R}=m^* \gamma_R$ and $k_F=\sqrt{2m^*\epsilon_F}$ to calculate the mediated exchange coupling in different regimes. For ballistic carriers, $\eta\rightarrow0$ is assumed in all calculations. For realistic materials or in the weak Rashba regime, we choose the spin splitting $\Delta_R=2k_F \gamma_R$= 5 meV and the Fermi energy $\epsilon_F$= 60 meV. Or equivalently, it corresponds to the Rashba coupling $\gamma_R =8.91\times10^{-12}$ eVm and the carrier density $n_{2D}=1.25\times10^{12}$ cm$^{-2}$. 

\begin{figure}
\centering
\includegraphics[width=7.6cm]{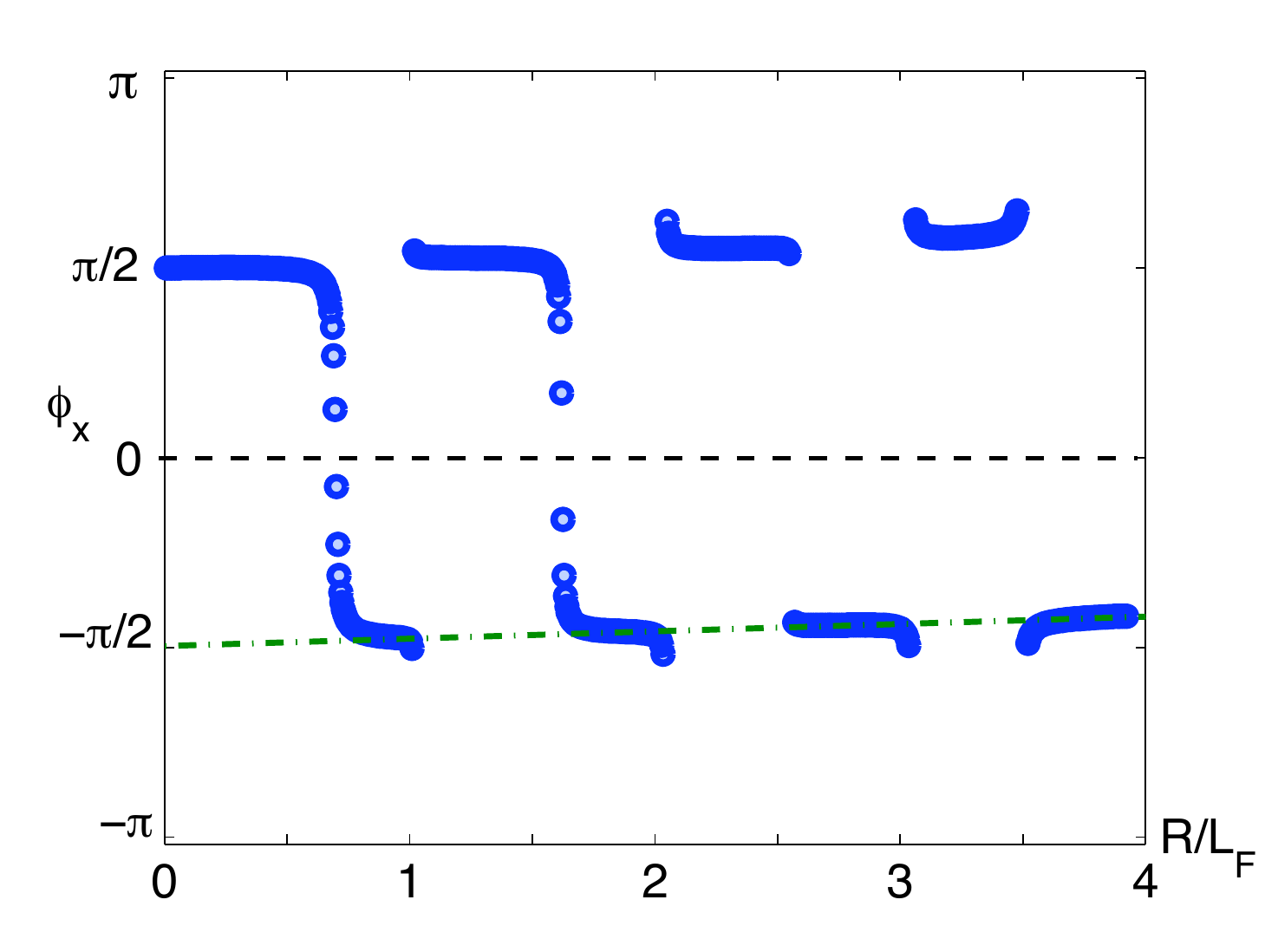}
\caption{The non-collinear angle $\phi_x$ in the weak Reshba regime with$k_{R}/k_{F}=0.02<<1$, $\eta\rightarrow0$ and $L_F$ is defined as $\frac{\pi}{k_F}$. The green dotted line is represented the upward tendency resulted from the Rashba effect. Notice that the lumps at discontinuous points are come from numerical error.} \label{RKKY} 
\end{figure}

The results are shown in Fig. \ref{RKKY}. In the weak Rashba regime $k_R/k_F\ll1$, the spatial trends of the angle $\phi_x(r)$ oscillates near the vicinity of $\pm \pi/2$ and is very similar to the ordinary RKKY interaction. The main difference is that, in the presence of the spin-orbit coupling, there is an upward background. The oscillatory behavior is characterized by $L_{F}\equiv 2\pi/2k_F$ as in ordinary RKKY oscillations. It is important to emphasize that the spatial trend is drastically different from the spin-precession argument in Datta-Das SFET\cite{Datta89} because the quantum interferences from different patches of the Fermi surface cannot be ignored.

The RKKY oscillation with an upward trend can be understood in a simple picture. Taking the asymptotic limit $k_F r \gg 1$, the reduced spin susceptibility along the radial direction $\chi_{ab}(r)$, where $a,b = x,z$, can be well approximated as 1D Rashba system. Applying a local gauge transformation\cite{Alei,Imamura04}, $U(r)=e^{-i k_R r \sigma^y/2}$, the Rashba Hamiltonian can be mapped into the 1D free electron gas with the well-known RKKY spin susceptibility. Since the local gauge transformation is nothing but the local rotation about the $y$-axis with the spiral angle $\phi(r) = k_R r$, the reduced susceptibility is approximately the usual RKKY oscillation twisted by a local spiral transformation,
\begin{equation}
\chi_{ab}(r) \approx \left[
\begin{array}{cc}
\cos k_R r & -\sin k_R r \\
\sin k_R r & \cos k_R r
\end{array} \right]_{ac} \chi^{RKKY}_{cb}(r),
\end{equation}
where the summation over the repeated index $c=x,z$ is implied. The gauge argument explains why our numerical results resemble the RKKY oscillation but with a gradual spiral background.

\begin{figure}
\centering
\includegraphics[width=7.6cm]{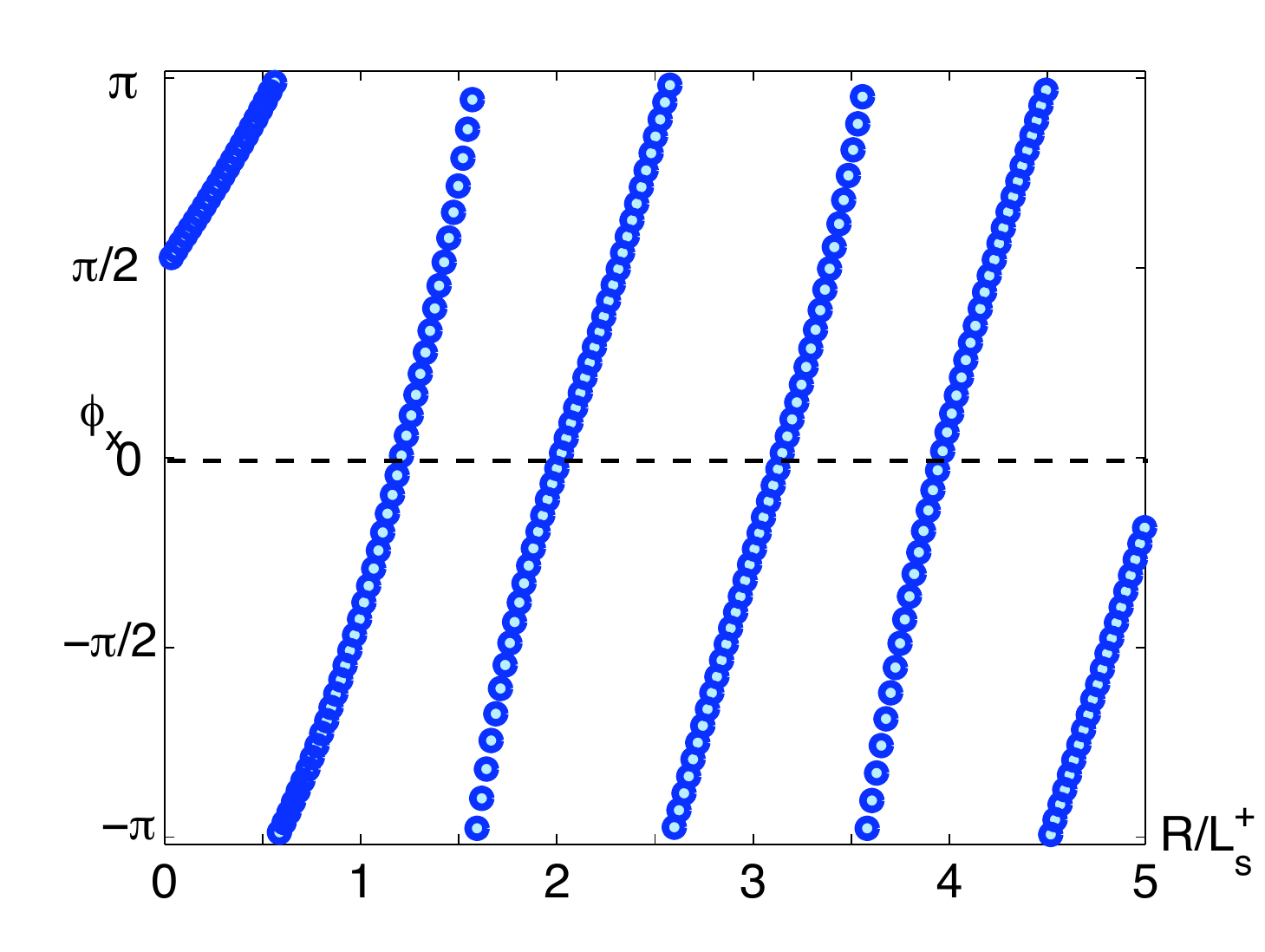}
\caption{The non-collinear angle $\phi_x$ in the dilute density regime with $k_{R}/k_{F}=3.5$ and $\eta\rightarrow0$. The period of the spiral angle is $L_s^+=\pi/(k_R+k_F)$.}\label{spiral}  
\end{figure}

By increasing the strength of Rashba coupling or lowering the electron density, we enter the strong coupling regime $k_R/k_F\gg1$. In this regime, the angle $\phi_x(r)$ ceases to oscillate and rotates gradually with the junction width as shown in Fig. \ref{spiral}. At first glance, the spiral evolution is similar to the spin-precession argument in Datta-Das SFET. But, it is not exactly so. The semiclassical spin precession is described by the length scale $L_s = 2\pi/(2 k_R)$. But, as Fig. \ref{spiral} shows, the spiraling length in our numerics is $L_{s}^{+}=2\pi/(2k_R+2k_F) < L_s$. Only in the extreme strong coupling where $k_F/k_R \ll 1$, the semiclassical picture will be completely correct.

It is important to mention that the momentum $2k_R+2k_F$ connects patches of the outer Fermi surface which are related by time-reversal symmetry.The period of the spiral angle indicates that the exchange coupling mainly contribute by the interference of the time-reversal section of outside chirality. Since the mediated exchange coupling is spiral-like, many phenomena, in this regime, related to the RKKY exchange coupling should be reconsidered. For example, a tunable non-local spin control in a coupled quantum dot system via a gate voltage has been beautifully demonstrated in the 2D electron gas system of semiconductor\cite{Craig04,Simon05,Simonin06}. The authors showed that the Kondo effect is suppressed by a nonlocal RKKY-like interaction which can be used to control the quantum dot spin. If the mediated exchange coupling is no longer RKKY-like, it is interesting to see how the non-collinear coupling will reshape the idea.  

\section{Path Integral Formalism}

In this section, we derive the effective exchange coupling analytically. Taking the asymptotic limit, the spatial trends of the non-collinear exchange coupling become rather clear. Within the path integral formalism, the effective coupling is obtained by integrating out the itinerant carriers and retaining the action to the quadratic order. After some algebra (details can be found in Appendix A), the effective Hamiltonian can be written as a contour integral\cite{Abricosov,Dugaev94,Imamura04}
\begin{eqnarray}\label{effHamiltonian}
&&\nonumber \hspace{-0.5cm}H_{\rm eff }=\frac{-J^2}{4\pi}\int_{-\infty}^{\mu}d\omega\ \ \mbox{Im}\left\{{\rm tr}\left[\left(\vec{\sigma}\cdot\vec{S}_1\right)G(\textbf{R},\omega+i\eta)\right. \right. \\
&&\hspace{3cm}\left. \left. \left(\vec{\sigma}\cdot\vec{S}_2\right)G(-\textbf{R},\omega+i\eta)\right]\right\},
\end{eqnarray}
where $\mu$ is the chemical potential, $\textbf{R}=\textbf{R}_1 -\textbf{R}_2$ and $\eta$ represents an infinitesimal imaginary energy, and tr means a trace over the spin degrees of freedom. In 2D Rashba system, the retarded Green's function in momentum space takes the form, 
\begin{eqnarray}
\nonumber &&\hspace{-0.45cm}G(\textbf{k};\epsilon+i0^+)=\left\{\epsilon+i0^+\hspace{-0.1cm}-\hspace{-0.1cm}\left[\frac{k^2}{2m^*}+\gamma_Rk\left(\hat{k}\times\hat{z}\right)\cdot\vec{\sigma}\right]\right\}^{-1}.\\ 
\end{eqnarray}
Fourier transforming it back to the coordinate space\cite{Dugaev94,Imamura04}, the retarded Green's function along the $x$-direction is
\begin{eqnarray}
G(R;\epsilon+i0^+)=G_0(R;\epsilon)\mathbf{1}+G_1(R;\epsilon)\sigma_y
\end{eqnarray}
with $G_0$ and $G_1$ defined as
\begin{eqnarray}
\nonumber &&\hspace{-0.5cm}G_0(R;\epsilon)=-i\frac{m^*}{4} \left\{\left(1+\frac{k_R}{q}\right) H_0^{(1)}\left[(q+k_R )R+i0^+\right]\right.\\&&\hspace{1.3cm}\left.+\left(1-\frac{k_R}{q}\right)H_0^{(1)}\left[(q-k_R)R+i0^+\right] \right\},\label{green1}\\
\nonumber &&\hspace{-0.5cm}G_1(R;\epsilon)=\frac{m^*}{4} \left\{\left(1+\frac{k_R}{q}\right) H_1^{(1)}\left[(q+k_R )R+i0^+\right]\right.\\&&\hspace{1.3cm}\left.-\left(1-\frac{k_R}{q}\right)H_1^{(1)}\left[(q-k_R)R+i0^+\right] \right\}\label{green2},
\end{eqnarray}
where $R=|\textbf{R}|$ and $q\equiv \sqrt{k_{\epsilon}^2 + k_R^2}$ with $k_{\epsilon}^2=2m^*\epsilon$. $H_n^{(1)}$ is the Hankel function of the first kind, defined as $H_n^{(1)}(z)=J_n(z)+iY_n(z)$ with $J_n$ and $Y_n$ being the Bessel functions of the first kind and second kind. Notice that the Bessel or Hankel functions have a branch cut along the negative real axis and the infinitesimal $i0^+$ is important in the spiral regime. We substitute the retarded Green's function into Eq.~(\ref{effHamiltonian}) and obtain the effective Hamiltonian along the $x$-axis, 
\begin{equation}
\mathcal{H}_{\rm eff}= A\vec{S_{1}} \cdot \vec{S_{2}} + B (\vec{S_{1}}\times \vec{S_{2}})_{y} + CS_1^y S_2^y,
\end{equation}
where coefficients $A$, $B$, and $C$ are
\begin{eqnarray}
&&A(R)=\frac{-J^2}{2\pi} \int_{\infty}^{\mu}d\epsilon\hspace{0.2cm}{\rm Im}\left[G_0^2 + G_1^2\right],\\
&&B(R)=\frac{J^2}{2\pi} \int_{-\infty}^{\mu}d\epsilon\hspace{0.2cm}2\hspace{0.1cm}{\rm Re}\left[G_0G_1\right],\\
&&C(R)=\frac{J^2}{2\pi} \int_{-\infty}^{\mu}d\epsilon \hspace{0.2cm}2\hspace{0.1cm}{\rm Re}\left[G_1^2\right].
\end{eqnarray}
The hard magnet is aligned along the $z$-axis, $\vec{S}=S_1^z\hat{z}$, so that we can rewrite the effective Hamiltonian as $\mathcal{H}_{\rm eff}= AS_1^z S_2^z + B S_1^z S_2^x$. The function $A(R)$ represents the collinear exchange coupling and $B(R)$ for the non-collinear part. By computing these two functions, the spiral angle can be determined.

Let us start with the weak Rashba coupling. In this regime, we have $q > k_R$, and by ignoring the slight contribution of $k_{\epsilon}<0$, the Green's functions can be simplified to be
\begin{eqnarray}
&& \hspace{-0.9cm}G_0
\simeq -\frac{im^*}{4}\hspace{-0.1cm}\left[H_0^{(1)}(qR+k_R R)+H_0^{(1)} (qR-k_R R)\right],\label{approxgreen1}\\
&&  \hspace{-0.9cm}G_1\simeq\frac{m^*}{4} \left[H_1^{(1)} (qR+k_R R) - H_1^{(1)} (qR-k_R R)\right].\label{approxgreen2}
\end{eqnarray}
With the asymptotic form of the Hankel functions,  
\begin{equation}
H_n^{(1)}(x)\simeq \sqrt{\frac{2}{\pi x}}\exp^{i\left(x-\frac{n\pi}{2}-\frac{\pi}{4}\right)},
\end{equation}
we end up with 
\begin{eqnarray}
&&G_0\simeq -i\frac{m^*}{\sqrt{2\pi q R}} \hspace{0.2cm}e^{i\left(qR-\frac{\pi}{4}\right)} \cos(k_{R}R),\\
&&G_1\simeq \frac{m^*}{\sqrt{2\pi q R}} \hspace{0.2cm}e^{i\left(qR-\frac{\pi}{4}\right)} \sin(k_{R}R).
\end{eqnarray}

The effective Hamiltonian is thus greatly simplified in the asymptotic limit,
\begin{eqnarray}\label{effRKKY}
\hspace{-0.2cm}\mathcal{H}^{\rm RKKY}_{\rm eff}\hspace{-0.1cm}\simeq \hspace{-0.1cm}F_1(R)\left[\cos(2k_RR)S_1^z S_2^z + \sin(2k_RR) S_1^z S_2^x\right]\hspace{-0.05cm},
\end{eqnarray}
where the range function\cite{Imamura04}
\begin{eqnarray}
F_1(R)=-\frac{J^2m^*}{8\pi^2}\frac{\sin(2q_FR)}{R^2}.
\end{eqnarray}
Note that the range function is identically the same as the usual two-dimensional RKKY range function\cite{Litvinov98}, except that $k_F$ is replaced by $q_F=\sqrt{2m^*\epsilon_F+k_R^2}$. Therefore, as clearly demonstrated in Eq.~(\ref{effRKKY}), the mediated exchange coupling is similar to the usual RKKY with a non-collinear spiral background as obtained numerically in Fig. \ref{RKKY}.

Now we turn to the strong coupling regime. In this regime, because $(q-k_R)R<0$, we need to be cautious about the analytic properties of the Hankel function $H_n^{(1)}(z)$. On the complex plane, $H_n^{(1)}(z)$ has a branch cut on the negative real axis. Thus, we need to pay extra attention to the $i0^+$ factor. Note that the Hankel function of the first kind along the negative real axis can be connected to the Hankel function of the second kind,
\begin{eqnarray}
H_n^{(1)}(-x+i0^+)=(-1)^{n+1} H_n^{(2)}(x+i0^+),
\end{eqnarray}
with the definition $H_n^{(2)}(z)=J_n(z)-iY_n(z)$. By this connection, we can rewrite the Eq.~(\ref{green1}) and (\ref{green2}) as,
\begin{eqnarray}
&& \hspace{-0.9cm}G_0\hspace{-0.1cm}
\simeq\hspace{-0.1cm} -\frac{im^*}{4}\frac{k_R}{q}\hspace{-0.1cm}\left[H_0^{(1)}(qR+k_R R)\hspace{-0.1cm}+\hspace{-0.1cm}H_0^{(2)} (k_R R-qR)\right],\\
&&  \hspace{-0.9cm}G_1\hspace{-0.1cm}\simeq\hspace{-0.1cm}\frac{m}{4} \frac{k_R}{q}\left[H_1^{(1)} (qR+k_R R)\hspace{-0.1cm} - \hspace{-0.1cm}H_1^{(2)} (k_R R-qR)\right].
\end{eqnarray}
In the asymptotic limit, the second kind of Hankel functions are
\begin{equation}
H_n^{(2)}(x)\simeq \sqrt{\frac{2}{\pi x}} e^{-i\left(x-\frac{n\pi}{2}-\frac{\pi}{4}\right)}.
\end{equation}
Simple algebra leads to the following results,
\begin{eqnarray}
&&G_0\simeq -i\frac{m^*}{\sqrt{2\pi k_R R}} \left(\frac{k_R}{q}\right) e^{iqR} \cos\left(k_R R-\frac{\pi}{4}\right),\\
&&G_1\simeq \frac{m^*}{\sqrt{2\pi k_R R}}\left(\frac{k_R}{q}\right) e^{iqR} \sin\left(k_R R-\frac{\pi}{4}\right).
\end{eqnarray}
Finally, the effective Hamiltonian takes the form,
\begin{eqnarray}\label{effspiral}
&&\hspace{-1.cm}\nonumber \mathcal{H}^{\rm sprial}_{\rm eff}\simeq F_2(R)\left[\cos\left(2k_R R-\frac{\pi}{2}\right)S_1^z S_2^z \right.\\
&&\hspace{2cm}\left.+ \sin\left(2k_R R-\frac{\pi}{2}\right) S_1^z S_2^x\right],
\end{eqnarray}
with the range function
\begin{eqnarray}
F_2(R)&=& \frac{J^2m^*}{4\pi^2}\frac{k_R}{R}\int_{0}^{q_F}dq\hspace{0.1cm}\frac{\sin(2qR)}{q}\\ &=&\frac{J^2m^*}{4\pi^2}\frac{k_R}{R}\hspace{0.2cm} {\rm Si}(2q_FR),
\end{eqnarray}
where ${\rm Si}(x)=\int_0^x\frac{\sin t}{t}dt$ is the sine integral function and is positive for $x>0$. In the asymptotic limit, $Si(x)$ reaches the constant value $\pi/2$ with minor oscillations. Thus, the spatial trend of the mediated exchange coupling is dominated by the spiraling part with minor oscillations. This is drastically different from the results in the weak Rashba regime. Again, our analytic results agree with those shown in Fig. \ref{spiral} obtained by numerical methods. 

\begin{figure}
\centering
\includegraphics[width=7.6cm]{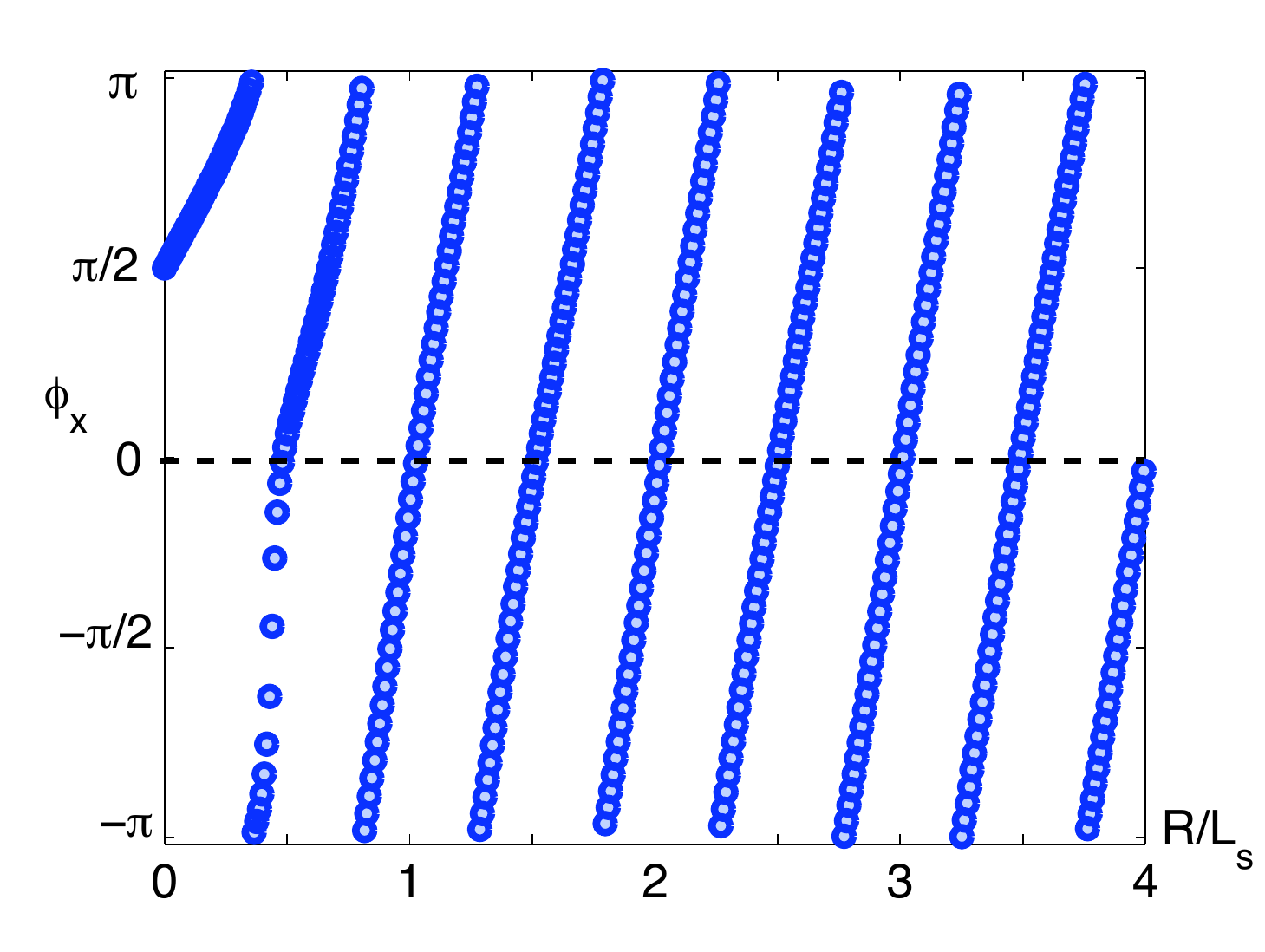}
\caption{The non-collinear angle $\phi_x$ in the critical regime with $k_{R}/k_{F}=1$ and $\eta\rightarrow0$. The spiral length at the critical point is $\pi/2k_R=L_s/2$, only half of that in the strong coupling regime.}\label{critical}
\end{figure}

The same calculation can be done for the critical case between the weak and the strong coupling regimes where one of the Fermi surface disappears. For comparison, the numerical results are shown in Fig. \ref{critical}. It is clear that the trend belongs to the spiral type. However, pay special attention to the length scale of the spiral. In the strong coupling regime, the spiral length is roughly $L_s^+$. Right at the critical point ($k_R = k_F$), the spiral period is $\pi/(2k_R) = L_s/2$, which follows the rule in the strong coupling regime. This factor of two can be understood as the more general formula for the length scale $L_s^+ = \pi/(k_R+k_F)$ in the strong coupling limit and at the critical point. The analytic calculations can be done by substituting $q=k_F=k_R$ into Eq.~(\ref{green1}) and Eq.~(\ref{green2}). The asymptotic forms of the Green functions near the Fermi surface are
\begin{eqnarray}
&&G_0=G_1\simeq -i\frac{m}{2\sqrt{\pi k_R R}} e^{i\left(2k_R R-\frac{\pi}{4}\right)},
\end{eqnarray}
and the effective Hamiltonian is
\begin{eqnarray}\label{effcritical}
\mathcal{H}^{\rm c}_{\rm eff}\propto \left[\cos(4k_RR)S_1^z S_2^z + \sin(4k_RR) S_1^z S_2^x\right]\hspace{-0.05cm},
\end{eqnarray}
with the spiral period $2\pi/4k_R = L_s/2$, as we mentioned above.

\section{Finite Spin Relaxation}

So far, we have revealed the interesting transitions between weak, strong and critical regimes assuming the mediating carriers are perfectly ballistic. In realistic materials, the spin relaxation rate is finite due to many mechanisms. Thus, it is interesting to study how the spatial trends of the mediated exchange coupling changes when the carriers are no longer perfectly ballistic. To fully address this issue, one needs to rederive the effective exchange coupling using a modified single-particle propagator. While this is certainly an interesting direction for future studies, we peek into the problem by introducing a phenomenological spin relaxation ration within the linear response theory. It is motivated by the observation that the inclusion of finite spin relaxation breaks time-reversal symmetry and shall suppress the RKKY interaction. Therefore, we expect the non-collinear spiral interaction should be enhanced and becomes dominant.

\begin{figure}
\centering
\includegraphics[width=7.6cm]{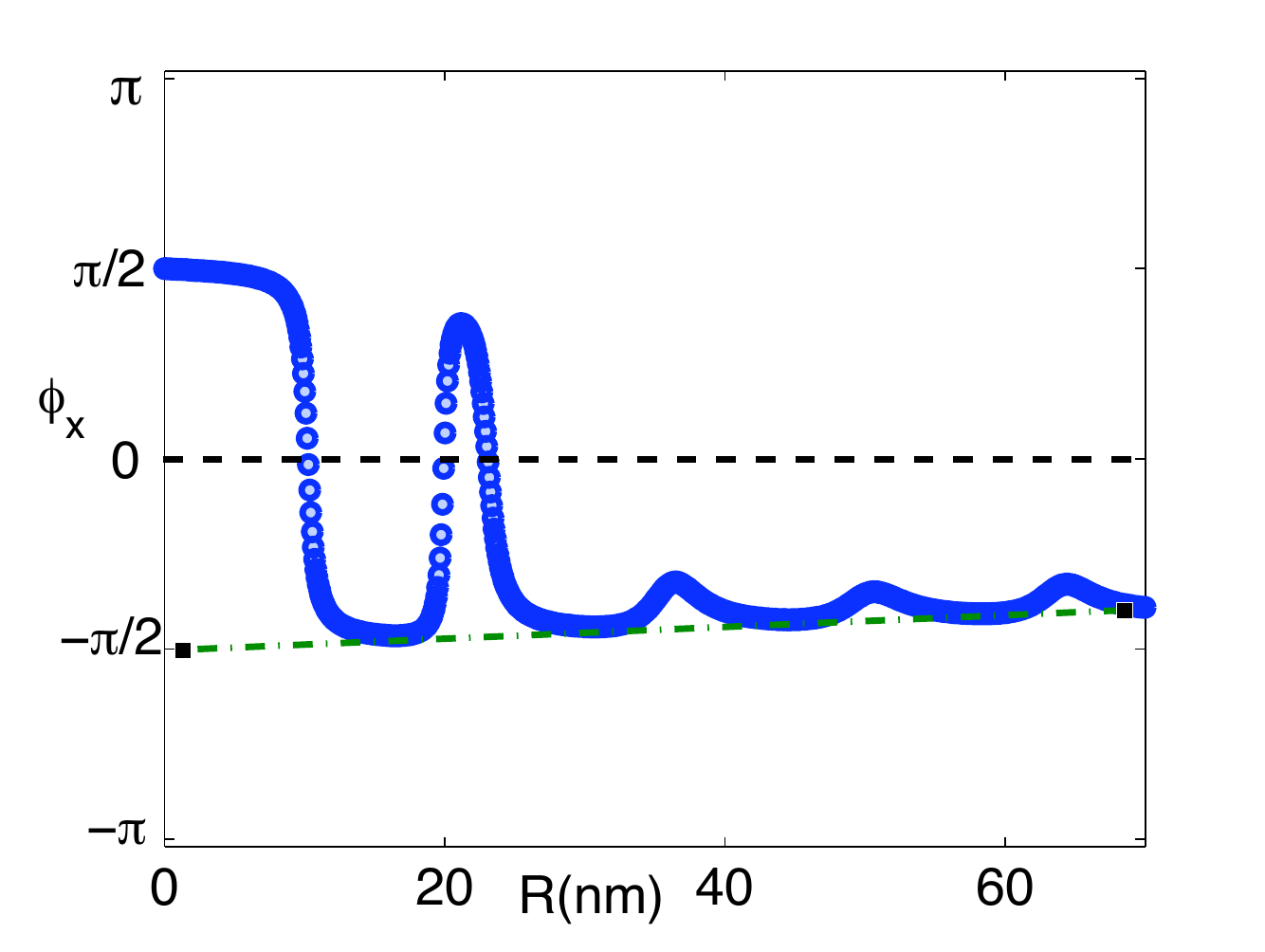}
\caption{Suppression of RKKY oscillations in the presence of spin relaxation. the RKKY phase with $k_R/k_F=0.02$ and $k_{\eta}/k_F=1$.}\label{relaxation1}
\end{figure}

\begin{figure}
\centering
\includegraphics[width=7.6cm]{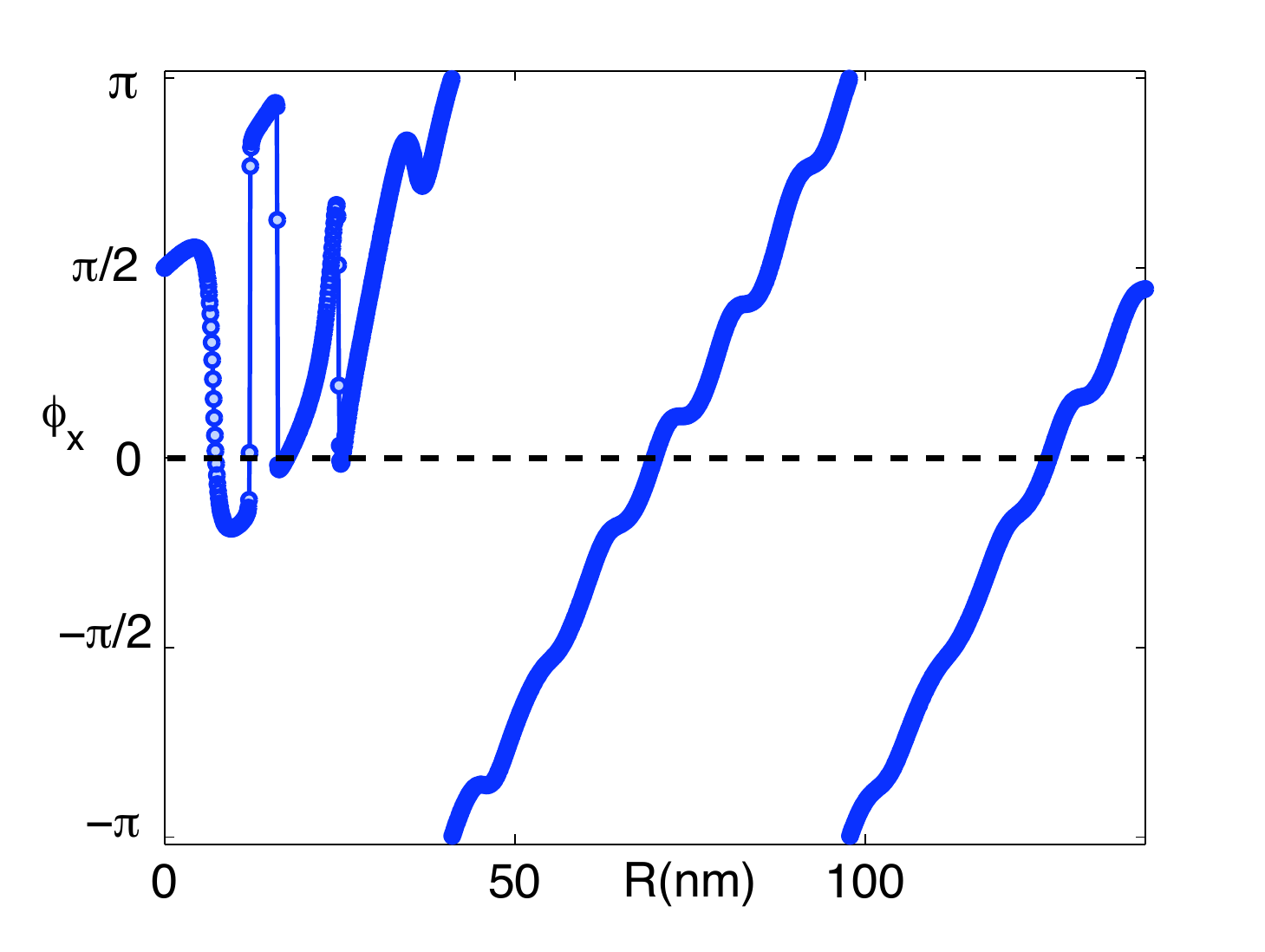}
\caption{Suppression of RKKY oscillations in the presence of spin relaxation with closer to the phase transition with $k_R/k_F=0.2$ and $k_{\eta}/k_F= 0.5$.}\label{relaxation2}
\end{figure}

For convenience, we introduce the parameter $k_{\eta}=\sqrt{2m^*\eta}$. In Fig. \ref{relaxation1}, we choose the parameters $k_{\eta}/k_F=1$ and $k_R/k_F=0.02$, and we can see that the RKKY pattern is indeed suppressed significantly due to the finite relaxation rate. Furthermore, as we move closer to the critical regime ( but still in the RKKY regime) with $k_R/k_F=0.2$, even a moderate spin relaxation $k_{\eta}/k_F=0.5$ will almost wash out the signature of RKKY oscillations and the spiral trend becomes rather robust, as shown in Fig \ref{relaxation2}. Therefore, it seems that the finite spin relaxation will suppress the RKKY oscillation and enhance the spiral interaction.

In conclusions, we apply both analytic and numeric methods to study the exchange coupling mediated by the itinerant carriers with Rashba interaction. We demonstrate that the Fermi surface topology greatly alters the property of mediated exchange coupling in 2D Rashba gas system. When the Rashba energy is much smaller than the Fermi energy, the mediated exchange coupling is almost the RKKY oscillation plus a slight upward trend. On the other hand, if the Rashba energy is larger than the Fermi energy, the mediated exchange coupling shows a spiral pattern. Apparently, the Fermi surface topology plays a crucial part about the mediated exchange coupling. We suggest that, in the quantum-dot systems \cite{Craig04,Simon05,Simonin06} with dilute carrier densities, the spiral-like exchange coupling should be considered instead of the usual RKKY interaction.

We acknowledge support from the National Science Council of Taiwan through grants NSC-96-2112-M-007-004 and NSC-97-2112-M-007-022-MY3 and also support from the National Center for Theoretical Sciences in Taiwan.

\appendix

\section{Derivations for Effective Hamiltonian}

Within the path integral approach, we investigate the effective coupling between two ferromagnets $\vec{S_1}$ and $\vec{S_2}$ mediated by itinerant carriers, as illustrated in Fig.\ref{junction}. Consider the Zener model, but only included two impurity spin, the Hamiltonian can be represented as
\begin{equation}
H=H_0+H_I,
\end{equation}
where
\begin{eqnarray}
H_0\hspace{-0.1cm}&=&\hspace{-0.2cm}\int d^D\textbf{r}\int d^D\textbf{r}' \sum_{\alpha}\Psi^{\dag}(\textbf{r})h_0(\textbf{r},\textbf{r}')\Psi(\textbf{r}'),\\
H_I\hspace{-0.1cm}&=&\hspace{-0.2cm}\int d^D\textbf{r}\hspace{-0.1cm}\sum_{I=1,2}\hspace{-0.1cm}\Psi^{\dag}(\textbf{r})\left[J\vec{S}_I (\textbf{r})\cdot\frac{\vec\sigma}{2}\right]\Psi(\textbf{r}').
\end{eqnarray} 
Here we ignore the correction effects of the itinerant carriers, and represent the unperturbed Hamiltonian $H_0$ as two fermionic operators. $\Psi$ is a two-component spinor for the itinerant carriers and $\vec{S}_I (\textbf{r}) =\vec{S}_I\delta(\textbf{r}-\textbf{R}_I)$ denotes the impurity spin at the position $\textbf{R}_I $ in the $D$-dimension, and $J$ represents the coupling between the itinerant and the localized spin densities.

In order to write down the partition function in path integral form, we introduce  the Grassmann variables $\phi$, $\bar{\phi}$ for the itinerant carriers\cite{Negele} and the coherent state for the impurity spins\cite{Auerbach}, which is denoted as $\vec{S}|\hat{\Omega}\rangle=S\hat{\Omega}|\hat{\Omega}\rangle$, where spin orientation is denoted here by $\hat{\Omega}=(\sin\theta\cos\psi,\sin\theta\sin\psi,\cos\theta)$  and the length of impurity spin vector by $S$. In path integral language, the partition function is represented as
\begin{equation}
\mathcal{Z}=\int\mathcal{D}[\vec{\Omega}_1,\vec{\Omega}_2]\int\mathcal{D}[\bar{\phi},\phi]\ \ e^{-\int_0^{\beta}d\tau L(\vec{\Omega}_1,\vec{\Omega}_2;\bar{\phi},\phi)},
\end{equation}   
where the integral denotes the summation of infinite time section. The Lagrangian can be writte as
\begin{eqnarray}
\nonumber \hspace{-0.3cm}L=&&\hspace{-0.3cm}\int d^D\textbf{r}\sum_{I=1,2}\langle \hat{\Omega}_I|\partial_{\tau}| \hat{\Omega}_I\rangle+\bar{\phi}(\textbf{r},\tau)\partial_{\tau}\phi(\textbf{r},\tau) \\ &&\hspace{3cm}+ H(\vec{\Omega}_1,\vec{\Omega}_2;\bar{\phi},\phi),
\end{eqnarray}  
where the first and second terms are the dynamics of the impurity spins and itinerant fermions respectively. Integrate out the itinerant carrier and arrive at an effective description for localized spin,
\begin{equation}
\mathcal{Z}=\int\mathcal{D}[\vec{\Omega}_1,\vec{\Omega}_2]\ \ e^{-\mathcal{S}_{\rm eff}(\vec{\Omega}_1,\vec{\Omega}_2)},
\end{equation}
with the effective action
\begin{eqnarray}
\mathcal{S}_{\rm eff}\hspace{-0.1cm}=\int_0^{\beta}\hspace{-0.1cm}d\tau\int d^D\textbf{r} \sum_{I=1,2}\langle \hat{\Omega}_I|\partial_{\tau}| \hat{\Omega}_I\rangle-\ln\left[\det\mathcal{G}^{-1}\right].
\end{eqnarray}
Here we use the notation,
\begin{eqnarray}
\mathcal{G}^{-1}(\vec{\Omega}_1,\vec{\Omega}_2)&=&\mathcal{G}_0^{-1}+\delta \mathcal{G}^{-1}(\vec{\Omega}_1,\vec{\Omega}_2),\\ 
\delta \mathcal{G}^{-1}(\vec{\Omega}_1,\vec{\Omega}_2)&=&\frac{J}{2}\left(\vec{\Omega}_1+\vec{\Omega}_2\right)\cdot\vec{\sigma}, \\ \mathcal{G}_0^{-1}&=&\partial_{\tau}\mathbf{1}-h_0.
\end{eqnarray}
By utilizing the relation, $\ln[\det M]={\rm tr}[\ln M]$, the second term of the effective action can be expanded, 
\begin{eqnarray}
\hspace{-0.7cm}\nonumber\ln\left[\det \mathcal{G}^{-1}\right]=&&\hspace{-0.3cm}{\rm tr}\left[\ln \left(\mathcal{G}_0^{-1}+\delta \mathcal{G}^{-1}\right)\right],\\ \nonumber=&&\hspace{-0.3cm}{\rm tr}\left[\ln \mathcal{G}_0^{-1}\right]+{\rm tr}\left[\ln\left(1+\mathcal{G}_0\delta \mathcal{G}^{-1}\right)\right],\\  =&&\hspace{-0.3cm}{\rm tr}\left[\ln\mathcal{G}_0^{-1}\right]-{\rm tr}\sum_{n=1}^{\infty}\frac{1}{n}\left(-\mathcal{G}_0\delta \mathcal{G}^{-1}\right)^n.
\end{eqnarray}
The zeroth order term does not contain any spin operators and can be neglected when computing the effective spin-spin interaction. The first-oder term is linear in the spin operator and vanishes since the itinerant carriers consider here are not polarized,
\begin{eqnarray}
\nonumber {\rm tr}\left[\mathcal{G}_0\delta \mathcal{G}^{-1}\right]&&\hspace{-0.3cm} =\frac{J}{2}\int d\tau \int d^D \textbf{r} \hspace{0.3cm}\mathcal{G}_0(\textbf{r},\tau;\textbf{r},\tau+0^{+})\\
&& \times \sum_{I=1,2}{\rm tr}\left\{\vec{\sigma}\cdot\vec{\Omega}_I(\textbf{r},\tau)\right\}=0,
\end{eqnarray}
where $\tau+0^+$ is for the creation operator. 

The second-order term contains two spin operators and is relevant for the effective exchange coupling,
\begin{eqnarray}
\nonumber &&\hspace{-0.45cm}S'_{\rm eff}\equiv\frac{1}{2}{\rm tr}\left[\mathcal{G}_0\delta \mathcal{G}^{-1}\mathcal{G}_0\delta \mathcal{G}^{-1}\right]\\\nonumber&&\hspace{0.15cm}=\frac{J^2}{4}\int_0^{\beta}d\tau\int_0^{\beta}d\tau'\hspace{0.1cm}{\rm tr} \left\{ \left[ \vec{\sigma}\cdot\vec{\Omega}_1(\tau)\right]\mathcal{G}_0(\textbf{R}_1,\tau;\textbf{R}_2,\tau')\right.\\ 
&&\hspace{2cm} \times
\left. \left[ \vec{\sigma}\cdot\vec{\Omega}_2(\tau')\right] \mathcal{G}_0(\textbf{R}_2,\tau';\textbf{R}_1,\tau) \right\}.
\end{eqnarray}
By truncating the expansion to second order, the effective action for the two impurity spins is
$\mathcal{S}_{\rm eff}= \int_0^{\beta}\hspace{-0.1cm}d\tau\int d^D\textbf{r} \sum_{I=1,2}\langle \hat{\Omega}_I|\partial_{\tau}| \hat{\Omega}_I\rangle+S'_{\rm eff}$. 
The first term of this effective action, known as Berry phase, describes the dynamics of the impurity spins. Here we assume that the dynamics of the itinerant spins is much faster so that we can ignore the retardation effect between two impurity spins. In other words, we take the stationary limit of the impurity spins $\vec{\Omega}_I(\tau)=\vec{S}_I$ and ignore the Berry-phase term. 

Now we would like to write down the effective action in the Fourier space of the imaginary time\cite{Flensberg},
\begin{eqnarray}
\mathcal{G}_0(\tau-\tau')=\frac{1}{\beta}\sum_{ik_n} \hspace{0.1cm} e^{-ik_n(\tau-\tau')}\mathcal{G}_0(ik_n)
\end{eqnarray}
where the summation is over all Matsubara frequencies $k_n=\frac{(2n+1)\pi}{\beta}$ for fermions. Making use of the identity
\begin{eqnarray}
\frac{1}{\beta}\int_0^{\beta}d\tau \ \ e^{i(k_m-k_n)\tau}=\delta_{mn}, 
\end{eqnarray}
the effective action after Fourier transformation is
\begin{eqnarray}\label{effective}
\nonumber  \hspace{-2cm}\mathcal{S}_{\rm eff}&=&\frac{J^2}{4}\sum_{ik_n}{\rm tr}  \left\{ \left[ \vec{\sigma}\cdot\vec{S}_1\right]\mathcal{G}_0(\textbf{R}_1,\textbf{R}_2;ik_n)\right.\\ 
&&\hspace{2cm}
\left. \left[ \vec{\sigma}\cdot\vec{S}_2\right] \mathcal{G}_0(\textbf{R}_2,\textbf{R}_1;ik_n) \right\}.
\end{eqnarray}
Notice that by neglecting the retardation effect, the poles of the Green's function, $\mathcal{G}_0(z)$, only show up on the real axis. This property let Eq.~(\ref{effective}) to be rewritten in a retarded Green's function form. 

First, we can demonstrate the stationary RKKY effect between two impurity spins mediated by free fermions from Eq.~(\ref{effective}). By Fourier transforming Eq.~(\ref{effective}) to the momentum space and fixing the direction of one impurity spin to the $z$-axis, we have ${\rm tr}\left[ \left(\vec{\sigma}\cdot S^z_1\hat{z}\right)\left(\vec{\sigma}\cdot \vec{S}_2\right)\right]=2S_1^zS_2^z$, and thus the equation can be rewritten as
\begin{eqnarray}\label{effaction}
\mathcal{S}_{\rm eff}&&\hspace{-0.3cm}=\int_{0}^{\beta}d\tau \hspace{0.2cm}H_{\rm eff},
\end{eqnarray}
with $H_{\rm eff}=J(\textbf{R}_{12})\hspace{0.1cm}S_1^zS_2^z$ and
\begin{eqnarray}
&&\nonumber \hspace{-1.cm}J(\textbf{R}_{12})=\frac{J^2}{2}\int\frac{d^D\textbf{q}}{(2\pi)^D}\int\frac{d^D\textbf{k}}{(2\pi)^D}\hspace{0.2cm}e^{i\textbf{q}\cdot\textbf{R}_{12}} \\ &&\hspace{1.2cm}\frac{1}{\beta}\sum_{ik_n}\mathcal{G}_0(\textbf{k}+\textbf{q};ik_n)\mathcal{G}_0(\textbf{k};ik_n).
\end{eqnarray}
Here we use the translation invariant property of free carriers and assume no spin polarization of this system. The summation of Matsubara frequencies for fermions can be computed and represented as the particle-hole propagation. After some calculation, we end up with 
\begin{eqnarray}\label{RKKYcoupling}
J(\textbf{R}_{12})=\frac{J^2}{2}\int\frac{d^D\textbf{q}}{(2\pi)^D}\hspace{0.2cm}e^{i\textbf{q}\cdot\textbf{R}_{12}} \hspace{0.2cm}\chi(\textbf{q}),	
\end{eqnarray}
where the particle-hole propagation is denoted as
\begin{eqnarray}
\chi(\textbf{q})=\int\frac{d^D\textbf{k}}{(2\pi)^D}\hspace{0.2cm}\frac{n_F(\xi_{\textbf{k}+\textbf{q}})-n_F(\xi_{\textbf{k}})}{\xi_{\textbf{k}}-\xi_{\textbf{k}+\textbf{q}}+i\eta},
\end{eqnarray}
where $\xi_{\textbf{k}}$ is the eigenenergy of free fermions and $n_F(\xi_{\textbf{k}})$ is the Fermi-Dirac distribution function. As shown as the Eq.~(\ref{RKKYcoupling}), we obtain the mediated RKKY exchange coupling strength between two impurity spins. 

\begin{figure}
\begin{center}
\includegraphics[width=7.1cm]{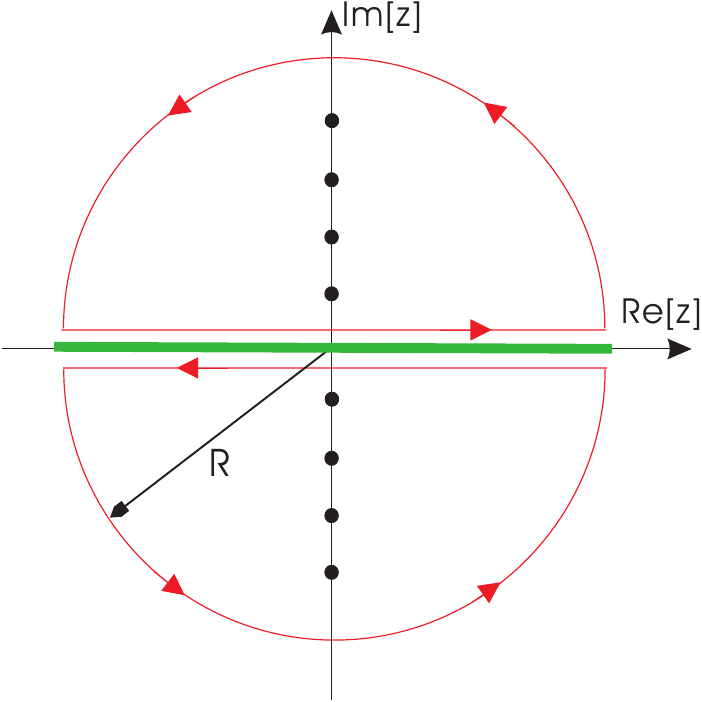}
\caption{Contours in the complex plane. The black points denotes the Matsubara frequencies $z=i\frac{(2n+1)\pi}{\beta}$ for fermions.}\label{contour}
\end{center}
\end{figure}

Next, we will replace  the imaginary Green's functions in Eq.~(\ref{effective}) by the retarded ones. The connection between these two types of Green's functions can be seen in the complex plane\cite{Flensberg}. The kernel of the Eq.~(\ref{effective}) in the complex plane, 
\begin{eqnarray}
\nonumber f(z)\hspace{-0.3cm}
&&={\rm tr}  \left\{ \left[ \vec{\sigma}\cdot\vec{S}_1\right]\mathcal{G}_0(\textbf{R}_1,\textbf{R}_2;z)\right.\\ 
&&\hspace{2cm}
\left. \left[ \vec{\sigma}\cdot\vec{S}_2\right] \mathcal{G}_0(\textbf{R}_2,\textbf{R}_1;z) \right\},
\end{eqnarray} 
has poles on the real axis. If we consider a contour integral,
\begin{eqnarray}
I=\int_{\mathcal{C}}\frac{dz}{2\pi i}\hspace{0.1cm}f(z)\hspace{0.1cm}n_F(z)\hspace{0.1cm}e^{z0^+},
\end{eqnarray}
where the contour path is shown in the Fig.\ref{contour}, it includes two parts, the upper and lower of the complex plane. The contour integral can be rewritten,
\begin{eqnarray}\label{imaginary}
I=\raisebox{-0.21cm}{\shortstack{Res\\ \scriptsize{\hspace{0.15cm}$z=ik_n$}}}\left[f(z)n_F(z)e^{z0^+}\right] =-\frac{1}{\beta}\sum_{ik_n}f(ik_n).
\end{eqnarray}
Here the Fermi-Dirac distribution is represent as $n_F(z)=(e^{\beta z}+1)^{-1}$, and has poles at $z=i(2n+1)\pi/\beta$, $n=\cdots,-1,0,1,\cdots$, where the residues at these poles are
\begin{eqnarray}
\hspace{-0.5cm}\raisebox{-0.21cm}{\shortstack{Res\\ \scriptsize{\hspace{0.15cm}$z=ik_n$}}}\left[n_F(z)\right]=\lim_{z\rightarrow ik_n}\frac{z-ik_n}{e^{\beta z}+1}=-\frac{1}{\beta}.
\end{eqnarray}
Because in the $|\textbf{R}|\rightarrow\infty$ limit, the factor $n_F(z)e^{z0^+}$ approaches zero whether ${\rm Re}(z)>0$ or not, the contour integral can be simplified to two line integrals very close to the real axis,
\begin{eqnarray}
&&\hspace{-1.3cm}\nonumber I=\lim_{\eta\rightarrow0^+}\left\{\int^{\infty+i\eta}_{-\infty+i\eta}\frac{dz}{2\pi i}n_F(z)f(z)\right.\\ &&\hspace{1.8cm}\left.+\int^{-\infty-i\eta}_{\infty-i\eta}\frac{dz}{2\pi i}n_F(z)f(z) \right\},\\&&\hspace{-1cm}=\lim_{\eta\rightarrow0^+}\int^{\infty}_{-\infty}\frac{d\omega}{2\pi i}n_F(\omega)\left[f(\omega+i\eta)-f(\omega-i\eta)\right].\label{retarded}
\end{eqnarray}
Combining Eq.~(\ref{imaginary}) and Eq.~(\ref{retarded}), we obtain the useful identity,
\begin{eqnarray}
&&\nonumber\hspace{-1.2cm}\sum_{ik_n}f(ik_n)=-\int_0^{\beta}d\tau\int^{\infty}_{-\infty}\frac{d\omega}{2\pi i}\hspace{0.2cm}n_F(\omega)\\ &&\hspace{2cm}\left[f(\omega+i\eta)-f(\omega-i\eta)\right],
\end{eqnarray}
where the first term of the right hand side is represented by the retarded Green's function and the second term is denoted the advanced Green's function. By using  the identity,
\begin{equation}
{\rm tr}\left[ABCD\right]^*={\rm tr}\left[B^{\dag}A^{\dag}D^{\dag}C^{\dag}\right]={\rm tr}\left[A^{\dag}D^{\dag}C^{\dag}B^{\dag}\right],\hspace{0.7cm}
\end{equation}
and the relation between advanced and retarded Green's function, 
$\left[G_0(\textbf{R}_2,\textbf{R}_1;\omega+i\eta)\right]^*=G_0(\textbf{R}_1,\textbf{R}_2;\omega-i\eta)$, we can obtain the second term as complex conjugate of the first. After reshaping the effective action, we end up with the effective Hamiltonian of the Eq.~(\ref{effaction}),
\begin{eqnarray}
&&\nonumber \hspace{-1.25cm}H_{\rm eff }=\frac{-J^2}{4\pi}\int_{-\infty}^{\infty}d\omega\hspace{0.1cm}n_F(\omega)\hspace{0.1cm} \textbf{Im}\left\{{\rm tr}\left[\left(\vec{\sigma}\cdot\vec{S}_1\right)\right. \right.\\ &&\left. \left.\hspace{-1.25cm}G_0(\textbf{R}_1,\textbf{R}_2;\omega+i\eta)\left(\vec{\sigma}\cdot\vec{S}_2\right)G_0(\textbf{R}_2,\textbf{R}_1;\omega+i\eta)\right]\right\}.
\end{eqnarray}
At $T=0$, the Fermi function is a step function. We can simply replace $n_F(\omega)$ by 1 and change the upper limit of the integral to the chemical potential $\mu$. Furthermore, if the unperturbed system is translational invariant, the effective Hamiltonian is just the function of the relative distance and we denote $\textbf{R}_{12}=\textbf{R}_1 -\textbf{R}_2$. In the end, the modified effective Hamiltonian comes to be the Eq.~(\ref{effHamiltonian}).


\begin{thebibliography}{999}

\bibitem{Wolf01}
S. A. Wolf, D. D. Awschalom, R. A. Buhrman, J. M. Daughton, S. von Moln\'ar, M. L. Roukes, A. Y. Chtchelkanova, and D. M. Treger,
Science {\bf 294}, 1488 (2001).

\bibitem{Zutic04}
I. Zutic, J. Fabian and S. Das Sarma,
Rev. Mod. Phys. {\bf 76}, 323 (2004).

\bibitem{Sun04}
S.-J. Sun and H.-H. Lin, 
Phys. Lett. A {\bf 327}, 73 (2004).


\bibitem{Sun06}
S.-J. Sun and H.-H. Lin, 
Eur. Phys. J. B {\bf 49}, 403 (2006).


\bibitem{Sharma}
P. Sharma, 
Science {\bf 307}, 531 (2005).


\bibitem{MacDonald05}
A. H. MacDonald, P. Schiffer and N. Samarth,
Nature Mat. {\bf 4}, 195 (2005).

\bibitem{Awschalom}
D. D. Awschalom, M. E. Flatt\'{e},
Nature Phys. {\bf 3}, 153 (2007).


\bibitem{Datta89}
S. Datta and B. Das, 
Appl. Phys. Lett. {\bf 56}, 665 (1989).

\bibitem{Rashba60}
E. I. Rashba,
Sov. Phys. Solid State {\bf 2}, 1109 (1960).





\bibitem{Francoeur}
S. Francoeur, M.-J. Seong, A. Mascarenhas, S. Tixier, M. Adamcyk and T. Tiedje,
Appl. Phys. Lett. {\bf 82}, 3874 (2003).

\bibitem{Fluegel}
B. Fluegel, S. Francoeur, A. Mascarenhas, S. Tixier, E. C. Young and T. Tiedje,
Phys. Rev. Lett. {\bf 97}, 067205 (2006).

\bibitem{Ast07}
C. R. Ast, J. Henk, A. Ernst, L. Moreschini, M. C. Falub, D. Pacil\'{e}, P. Bruno, K. Kern, and M. Grioni,
Phys. Rev. Lett. {\bf 98}, 186807 (2007).

\bibitem{Ast08}
C. R. Ast, D. Pacil\'{e}, L. Moreschini, M. C. Falub, M. Papagno, K. Kern, M. Grioni, J. Henk, A. Ernst, S. Ostanin and P. Bruno,
Phys. Rev. B {\bf 77}, 081407(R) (2008).




\bibitem{DP}
M. I. Dyakonov and V. I. Perel, 
Sov. Phys. Solid State {\bf 13}, 3023 (1972).

\bibitem{Grimaldi05}
C. Grimaldi, 
Phys. Rev. B {\bf 72}, 075307 (2005).

\bibitem{Rashba04}
E. I. Rashba, 
Phys. Rev. B {\bf 70}, 201309(R) (2004).

\bibitem{Dimitrova}
O. V. Dimitrova, 
Phys. Rev. B {\bf 71}, 245327 (2005).

\bibitem{Grimaldi06}
C. Grimaldi, E. Cappelluti and F. Marsiglio, 
Phys. Rev. B {\bf 73}, 081303(R) (2006).

\bibitem{Chaplik}
A. V. Chaplik and L. I. Magarill, 
Phys. Rev. Lett. {\bf 96}, 126402 (2006).

\bibitem{Galstyan}
A. G. Galstyan and M. E. Raikh, 
Phys. Rev. B {\bf 58}, 6736 (1998).

\bibitem{Grimaldi08}
C. Grimaldi, 
Phys. Rev. B {\bf 77}, 113308 (2008).

\bibitem{Cappelluti}
E. Cappelluti,C. Grimaldi and F. Marsiglio,
Phys. Rev. Lett. {\bf 98}, 167002 (2007).

\bibitem{Huang06}
W.-M. Huang, C.-H. Chang, and H.-H. Lin, 
Phys. Rev. B {\bf 73}, 241307(R) (2006).

\bibitem{Huang08}
W.-M. Huang, H.-H. Lai, C.-H. Chang, and H.-H. Lin, 
Int. J. Mod. Phys. B {\bf 22}, 88 (2008).

\bibitem{Imamura04}
H. Imamura, P. Bruno, and Y. Utsumi, 
Phys. Rev. B {\bf 69}, 121303(R) (2004).



\bibitem{Craig04}
N. J. Craig, J. M. Taylor, E. A. Lester, C. M. Marcus, 
M. P. Hanson, and A. C. Gossard,
Science {\bf 304}, 565 (2004).

\bibitem{Simon05}
P. Simon, R. L\'{o}pez, and Y. Oreg , 
Phys. Rev. Lett. {\bf 94}, 086602 (2005).

\bibitem{Simonin06}
J. Simonin, 
Phys. Rev. Lett. {\bf 97}, 266804 (2006).



\bibitem{Lin04}
S.-J. Sun, S.-S. Cheng and H.-H. Lin, 
Appl. Phys. Lett. {\bf 84}, 2862 (2004).
 
\bibitem{Lin06}
C.-H. Lin, H.-H. Lin and T.-M. Hong 
Appl. Phys. Lett. {\bf 89}, 032503 (2006).

\bibitem{Flensberg}
H. Bruus and K. Flensberg, 
{\em Many-Body Quantum Theory in Condensed Matter Physics -- An Introduction} (Oxford University Press, 2004).

\bibitem{Mahan}
G. D. Mahan, 
{\em Many-Particle Physics} (springer, 3nd edition, 2007).

\bibitem{Alei}
I.L. Aleiner and V.I. Fal'ko, 
Phys. Rev. Lett. {\bf 87}, 256801 (2001).




\bibitem{Abricosov}
A. A. Abricosov, L. P. Gorkov, and I. E. Dzyaloshinski, 
{\em Quantum Field Theoretical Method in Statistical Mechanics} (Pergamon, New York, 1965).

\bibitem{Dugaev94}
V.K. Dugaev, V.I. Litvinov, and P.P. Petrov, 
Superlattices Microstruct. {\bf 16}, 413 (1994).

\bibitem{Litvinov98}
V.I. Litvinov, and V. K. Dugaev, 
Phys. Rev. B {\bf 58}, 3584 (1998).




\bibitem{Negele}
J. W. Negele, H. Orland, 
{\em Quantum Many-particle Systems} (Westview Press, 1998)


\bibitem{Auerbach}
A. Auerbach, 
{\em Interacting Electrons and Quantum Magnetism} (Springer Verlag, 1994).




\end{thebibliography}
\end{document}